\documentclass{article}
\usepackage[colorlinks,linkcolor=blue,urlcolor=cyan]{hyperref} 
\usepackage{graphicx} 
\usepackage{multirow}
\usepackage{pdflscape}
\usepackage{caption}
\usepackage{biblatex} 
\usepackage[font=small,labelfont=bf]{caption} 
\graphicspath{ {images/} }

\title{Power Analysis for Experiments with Clustered Data, Ratio Metrics, and Regression for Covariate Adjustment}
\author{Tim Hesterberg and Benjamin Knight (Instacart)
  \thanks{We thank Mack Sweeney and Michael Curran for helpful comments.}
}
\date{June 10, 2024}

\newcommand{\eg}{e.g.\ }
\newcommand{\etc}{etc.\ }
\newcommand{\ie}{i.e.\ }

\newcommand\Var {\mathrm{Var}}

\newcommand\SE {\mathrm{SE}}
\newcommand\MDE {\mathrm{MDE}}

\newcommand\betahat {\hat\beta}
\newcommand\gammahat {\hat\gamma}
\newcommand\Deltahat {\hat\Delta}

\newcommand\thetahat {\hat\theta}

\newcommand\muhat {\hat\mu}

\newcommand\xbar {\bar x}

\newcommand\wbar {\bar w}
\newcommand\Wbar {\bar W}
\newcommand\ybar {\bar y}
\newcommand\Ybar {\bar Y}
\newcommand\yhat {\hat y}
\newcommand\Yhat {\hat Y}

\newcommand\what {\hat w}

\newcommand\sumin {\sum_{i=1}^n}
\newcommand\meanin {n^{-1} \sum_{i=1}^n}

\begin{document}
\maketitle


\section*{Abstract}

We describe how to calculate standard errors for A/B tests
that include clustered data, ratio metrics, and/or covariate adjustment.
We may do this
for power analysis/sample size calculations
prior to running an experiment using historical data,
or after an experiment for hypothesis testing and confidence intervals.
The different applications have a common framework, using
the sample variance of certain residuals.
The framework is compatible with modular software, can be
plugged into standard tools, doesn't require computing covariance matrices,
and is numerically stable.
Using this approach we estimate that covariate adjustment
gives a median 66\% variance reduction for a key metric,
reducing experiment run time by 66\%.

\paragraph{Keywords:}
A/B experiment, regression adjustment, CUPED, ANCOVA,
variance reduction, sample size calculation, standard error, delta method.


\section{Motivation}

\indent\indent
When running A/B tests (randomized controlled trials), time is money. The faster we can run experiments, the faster we can ship promising treatments. We might put a lot of effort into variance-reduction techniques to obtain more accurate answers, but if power analysis/sample size planning tools don't reflect that then the experiments we design will run longer than necessary.

Most off-the-shelf power analysis tools handle the simple case where
analysis uses $t$-tests for independent observations, but not clustered
data, ratio metrics, or variance-reduction methods.

In this article we present a framework for conducting power analysis for A/B tests that can support any combination of the following applications:
\begin{description}
  \item[Clustered Data:] If we are interested in testing a feature that improves the customer experience, then the most intuitive unit of randomization for an A/B test is the customer. However, customers may place multiple orders. If the metric of interest is at the level of the order (\eg mean order size) then we need to take this clustering into account when calculating standard errors.
  \item[Ratio Metrics:] Some metrics are a ratio between two random quantities,
  \eg `Revenue Share from Electronics' = (revenue from electronics)/(total revenue). Standard errors depend on the variances of the numerator, denominator, and their correlation.
  \item[Covariate Adjustment:] While random assignment makes experiment arms balanced on average, random imbalances do occur. We can reduce the variance of estimates by correcting for this covariate imbalance using regression. Standard errors should reflect this improvement.
\end{description}

These applications reduce to four basic cases, combinations of
simple means or ratio metrics (including clustered data), with or without covariate adjustment.
In all cases we obtain standard errors
using the sample standard deviations of certain residuals.
We begin with a review of power analysis for unadjusted means
in Section~\ref{section:refresher}, and consider the other cases
in Sections~\ref{section:ratioMetrics}, \ref{section:covariateAdjustment}
and \ref{section:ratioAdjustment}.
Section~\ref{section:applications} includes a summary
of the standard errors and residuals in
\hyperlink{summaryTable}{Table~\ref{table:SEsummmary}},
then includes an example and a meta-analysis from Instacart.

\section{A Refresher on Conventional Power Analysis}
\label{section:refresher}

\indent\indent Power analysis (or sample size planning) involves relationships between
four parameters of interest:
\begin{enumerate}
  \item sample size ($n$) representing the number of units selected for experimental assignment,
  \item false positive rate (type I error rate) $\alpha$,
  \item power ($1 - \beta$, where $\beta$ is the type II error rate), the probability of detecting differences of a given magnitude, and
  \item minimum detectable effect (MDE) --- the change in the response variable that is detectable with that power.
\end{enumerate}

For simplicity we focus on two-arm experiments (``control'' and
``treatment'' arms, denoted C and T) and focus on power and sample size
estimates for a single metric. Let $\Deltahat$ be an estimate of the
treatment effect, (\eg difference of means between T and C), and
$\SE_{\Deltahat}$ be its standard error.
We focus on one-sided tests, because the vast majority of experiments
at Instacart are run for the purpose of testing whether a treatment
causes a metric to improve.
We assume that sample sizes are large enough that
both estimates and their corresponding $t$-statistics
($\Deltahat / \SE_{\Deltahat}$) are approximately normally distributed.

The four parameters are related by the equation
\begin{equation}
\MDE_{\Deltahat}
  = (z_\alpha + z_\beta) \SE_{\Deltahat}
\end{equation}
where $z_\alpha$ and $z_\beta$ are the Normal quantiles corresponding to the type I and II error rates respectively,
and $\SE_{\Deltahat}$ depends on $n$.

\subsection{Difference of Means}

For the simple case of a difference in means assuming equal variances,
no clustering, and equal sample sizes
\begin{equation}
\MDE_{\Deltahat}
  = (z_\alpha + z_\beta) \frac{2s_y}{\sqrt{n}}
\label{equation:mdeTwoEqualMeans}
\end{equation}
where $s_y$ is the sample standard deviation of the response variable. Then given $s_y$ and any three of $\MDE$, $n$, $\alpha$ and $\beta$ we can calculate the fourth.

The factor of $2$ arises in a two-armed experiment. Suppose that there are
$n_C$ and $n_T$ observations in the C and T arms, with sample standard deviations
$s_C$ and $s_T$, then
\begin{equation}
  \SE_{\Ybar_T - \Ybar_C} = \sqrt{\frac{s^2_C}{n_C} + \frac{s^2_T}{n_T}}
  \label{equation:unpooled}
\end{equation}
But when planning an experiment we don't have those sample standard deviations or the actual sample sizes; instead we typically estimate both sample standard deviations using a single value $s_y$ estimated from historical data, and specify what fraction of observations $\psi$ will be allocated to the treatment group;
then Equation~(\ref{equation:unpooled}) reduces to
\begin{equation}
  \SE_{\Ybar_T - \Ybar_C} = \sqrt{\frac{s^2_Y}{(1-\psi) n} + \frac{s^2_Y}{\psi n}}
  = \sqrt{\frac{1}{1-\psi} + \frac{1}{\psi}} \frac{s_y}{\sqrt{n}}
\label{equation:seDiffMeans}
\end{equation}
In the special case that $\psi = 50\%$,
$\sqrt{\frac{1}{1-\psi} + \frac{1}{\psi}} = 2$,
as in Equation~(\ref{equation:mdeTwoEqualMeans}).

\begin{figure}[h]
\centering
\includegraphics[width=1.0\textwidth]{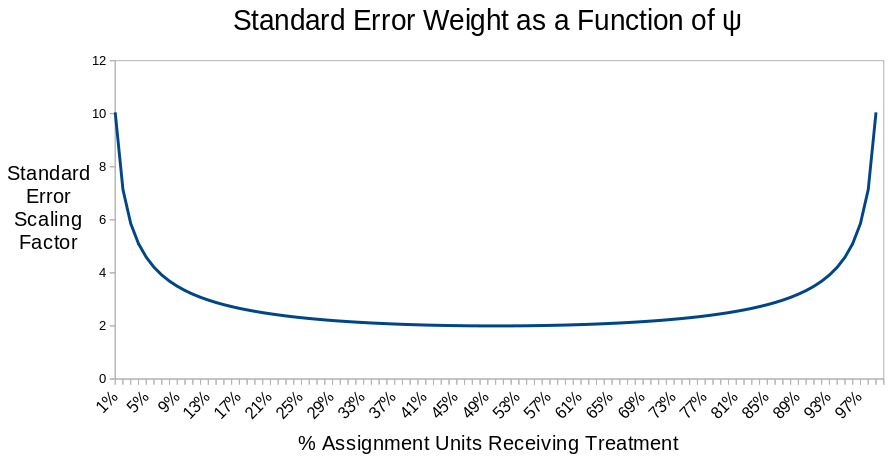}
\caption{Standard error multiplier $\sqrt{1/(1-\psi)+1/\psi}$. The standard error for the difference of means when one receives fraction $\psi$ of the total sample size $n$ is this factor times $s/\sqrt{n}$.
\label{figure:psiEffect}}
\end{figure}

More generally, Figure~\ref{figure:psiEffect} shows how standard errors
depend on $\psi$. The minimum scaling factor is $2$ at $50\%$, and is slightly
larger for values near $50\%$, but increases dramatically
when the fraction approaches $0$ or $1$.

Also note that $p$-values may be inaccurate if
metrics are skewed and the split is not 50-50.
The old ``$n \ge 30$'' rule for the Central Limit Theorem
is badly wrong for skewed data.
$p$-values from
a one-sample $t$ test are not
reasonably accurate until $n > 5000$ for an exponential
population\footnote{See \cite{hest15c}
for more about skewed data.
}
or $n \ge 125,000$ for some important skewed metrics at Instacart.
Two-sample tests with a 50-50 split are better because
the skewness cancels out for $\Ybar_T - \Ybar_C$.

\subsection{Summary for Difference of Means}

To recap, for a 50-50 split
\begin{equation}
\MDE_{\Deltahat} = (z_\alpha + z_\beta) \frac{2 s_y}{\sqrt{n}}
\label{equation:mdeDiffMeans}
\end{equation}

The sample size necessary to achieve a specified MDE is
\begin{equation}
  n \ge \left(\frac{(z_\alpha + z_\beta) 2 s_y}{\MDE}\right)^2
\label{equation:nDiffMeans}
\end{equation}

For splits other than 50-50, substitute
$\sqrt{\frac{1}{1-\psi} + \frac{1}{\psi}}$ for $2$,
but take care to check that skewness does not invalidate normal approximations.

\subsection{Generalizing Beyond Difference of Means}

\begin{figure}[ht]
\centering
\includegraphics[width=1.0\textwidth]{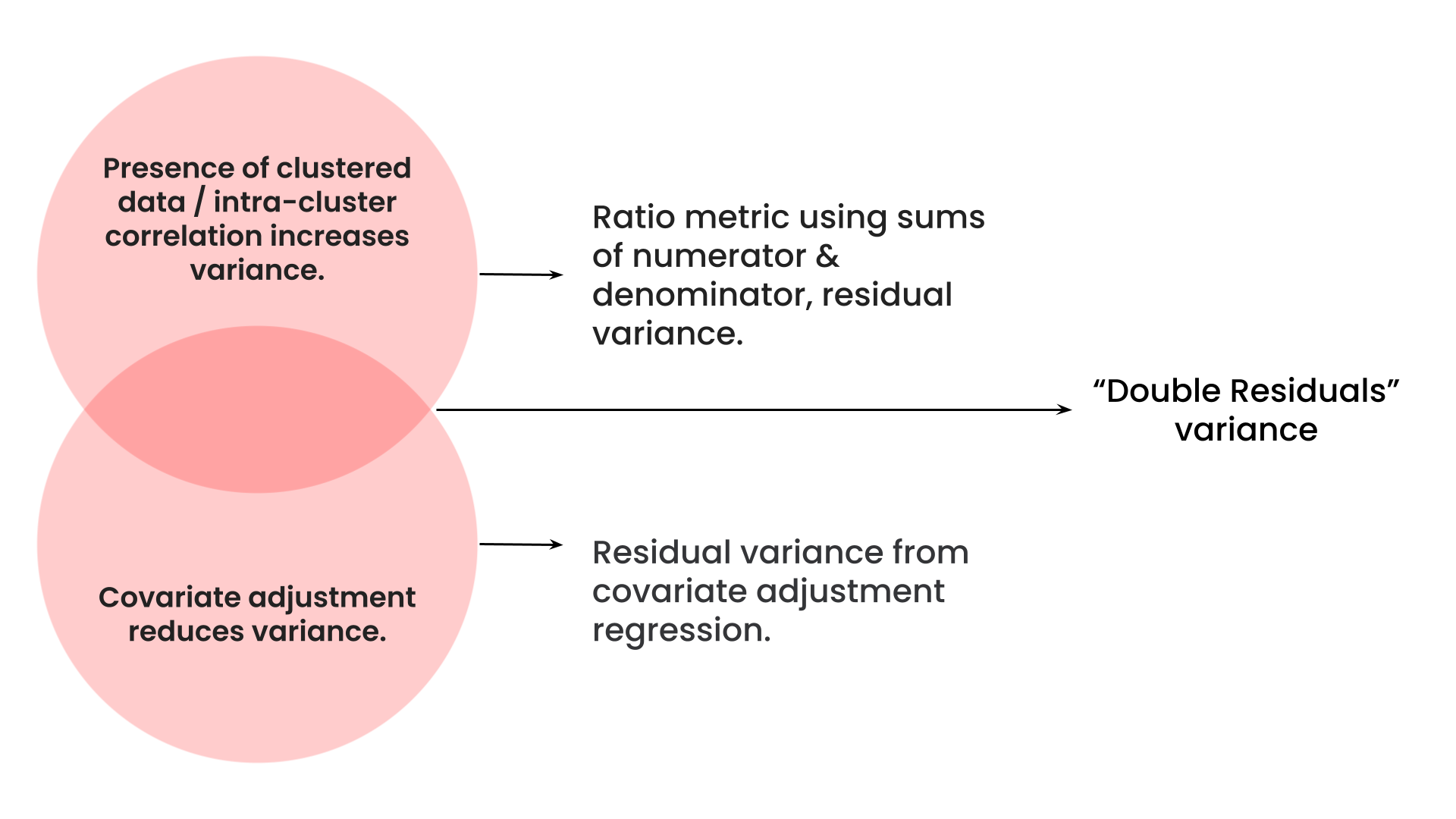}
\caption{Strategy for estimating standard errors given multiple observations per unit of experimental assignment, covariate adjustment, and their combination.
\label{figure:vennDiagram}}
\end{figure}

It turns out that
Equations~(\ref{equation:mdeDiffMeans}--\ref{equation:nDiffMeans})
almost work
for clustered data, ratio metrics, and covariate adjustment
applications --- we
just need to replace the value $s_y$ with other quantities that are
based on residual standard deviations.
Our broad strategy for deriving these values is shown in
Figure~\ref{figure:vennDiagram}.

To correctly estimate standard errors we need to account for two factors.
First, when data are clustered there is intra-cluster correlation (the top portion  of the Venn diagram);
ignoring this typically results in standard errors that are too small, causing inflated false positive rates and too-short confidence intervals.
Second, controlling for random imbalances in covariates between arms reduces the
variability of estimates; ignoring this results in
too-large standard errors.
Finally, these factors may occur together.
In subsequent sections we describe how to estimate standard errors
in these cases, using ratio estimates and residual standard
deviations.

\section{Ratio Metrics and  Clustered Data}
\label{section:ratioMetrics}

\indent\indent
In this section we discuss how ratio metrics arise, either due
to clustering or natural ratio metrics, and derive standard errors.

Our first challenge is clustered data.
For example, consider estimating average order size (in dollars).
We call this GMV per order (Gross Merchandise Value).
Calculating the metric is straightforward, as the total
value of items ordered divided by the number of orders.
Calculating the standard error is not.
We must account for correlations within clusters (for example, orders
created by the same customer will tend to be of similar sizes).

We begin by aggregating the data by cluster to obtain two
values for each customer: $Y_i =$ total value of items ordered
by customer $i$, and $n_i =$ number of orders by customer $i$.
Then the metric is a ratio of $\sum_i Y_i/\sum_i n_i$,
or equivalently the ratio of two sample means $\Ybar / \bar n$.
This simplifies the problem in one way ---
we now have independent observations --- but complicates
it in others. Instead of a sample mean, we have a ratio
of two sample means, and the numerator and denominator are dependent.

Other metrics represent naturally occurring ratios, even without clustering.
For example, some retailers have their own in-store workers pick some orders, then Instacart shoppers deliver them to customers.
The fraction of GMV picked by Instacart shoppers is a ratio:
GMV picked by Instacart shoppers / total GMV.

Clustering may also occur
with such natural ratio metrics, \eg clustering from the order level to
shopper or store level.

We use the following notation to handle ratio metrics, with or without
clustering.
$Y$ corresponds to the metric of interest, or numerator of a ratio.
Where there is clustering, we let
\[
Y_{ij} = \mbox{Cluster}_i \mbox{,\ Observation}_j \mbox{(\eg\phantom{a}customer\phantom{a}}i\mbox{,\phantom{a}order \phantom{a}}j)
\]
\[
  Y_i = \sum_j Y_{ij} = \mbox{Sum for Cluster\phantom{a}}i
\]
$W$ corresponds to the denominator, to a cluster size or count,
\[
  W_i = \sum_j W_{ij} {~or~} n_i
\]
The individual or cluster ratio is
\[
  V_i = \mbox{Cluster-level average (or ratio): } Y_i/W_i
\]
We estimate the metric or ratio of interest as:
\begin{equation}
  \thetahat = \frac{\sum_{ij} Y_{ij}}{\sum_{ij} W_{ij}} = \frac{\sum_i W_i V_i}{\sum_i W_i} = \frac{\sum_i Y_i}{\sum_i W_i} = \frac{\bar Y}{\bar W}
  \label{equation:ratioMetric}
\end{equation}

While it might be natural
to think of these metrics as weighted averages $\sum W_i Y_i / \sum W_i$,
that makes calculating standard errors tricky --- see the appendix.
Instead we estimate standard errors for ratio metrics using the delta method.

\subsection{Standard Errors Using the Delta Method and Residuals}

\begin{figure}[ht]
\centering
\includegraphics{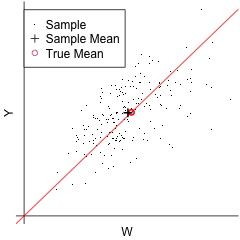}
\caption{
The true slope $\theta$ is the slope of a line through
the origin and the true mean $(\mu_W, \mu_Y)$ (shown in red).
The estimated slope $\thetahat$ is the slope of the line through
the origin and the sample mean $(\wbar, \ybar)$ (line not shown).
Even though $\Ybar < \mu_Y$, the estimated slope $\thetahat$ is
greater than the true slope $\theta$ because the average residual
is greater than zero.
Let $\bar R = \Ybar - \theta \Wbar$ be
the average residual relative to the true line.
The difference $\thetahat - \theta$ between the true and estimated slopes
is exactly equal to $\bar R / \Wbar$,
and is approximately equal to $\bar R / \mu_W$.
For large samples the error in the approximation is small, because
the difference between $\Wbar$ and $\mu_W$ is small, and $\bar R$ is small.
\label{figure:deltaMethodRatio}
}
\end{figure}

We turn now to calculating standard errors for ratio metrics,
whether due to clustering or not.
We use the
delta method.
We find a linear approximation to the
ratio, based on a first-order bivariate Taylor series
of the function $f(\Wbar, \Ybar) = \Ybar / \Wbar$ about $(\mu_W, \mu_Y)$,
\begin{eqnarray}
  \thetahat = \frac{\bar Y}{\bar W}
  &=& \frac{\mu_Y}{\mu_W} + \frac{\bar Y - (\mu_Y/\mu_W) \bar W}{\Wbar}
      \nonumber\\
  &\approx& \frac{\mu_Y}{\mu_W} + \frac{\bar Y - (\mu_Y/\mu_W) \bar W}{\mu_W}
  = \theta + \frac{\bar Y - \theta \bar W}{\mu_W}.
        \label{equation:deltaMethod}
\end{eqnarray}
where $\mu_Y$ and $\mu_W$ are the population means for the numerator and denominator, respectively. The estimate $\thetahat$ approximately equals the
true value of the ratio, plus the mean residual divided by the true mean
denominator.
We visualize this in Figure~\ref{figure:deltaMethodRatio}.

\clearpage

Then the variance approximation is
\begin{equation}
  \Var(\thetahat)
  \approx \frac{1}{\mu_W^2} \Var(\bar Y - \theta \bar W)
  = \frac{1}{n \mu_W^2} \Var(Y - \theta W)
  \label{equation:varianceRatioEstimate}
\end{equation}

A common next step would be to expand
$\Var(Y - \theta W)$
using variances and covariances.
We prefer not to do this.
Thinking of the variance in terms of the variance of residuals is
easier to understand, particularly as we consider covariate adjustment below.
Furthermore, that expansion can result in numerically-unstable
estimates, including negative variances.

To use Equation~(\ref{equation:varianceRatioEstimate}) in practice,
we substitute estimates for unknown quantities:
\begin{equation}
  \widehat{\Var}(\thetahat)
  = \frac{1}{n \wbar^2} \widehat{\Var}(Y - \thetahat W)
  = \frac{1}{n \wbar^2} s^2_r
\end{equation}
where
\begin{equation}
  s_r^2 = \frac{1}{n-1} \sumin r_i^2
\end{equation}
is the sample variance
of the empirical residuals
\begin{equation}
  r_i = y_i - \thetahat w_i.
  \label{equation:ratioResiduals}
\end{equation}
The standard error is
\begin{equation}
  \SE(\thetahat)
  = \frac{1}{\wbar} \frac{s_r}{\sqrt{n}}
  \label{equation:SEtheta}.
\end{equation}

\section{Covariate Adjustment}
\label{section:covariateAdjustment}

\indent\indent In a randomized controlled trial the assignment
of subjects to arms is fair on average, but in any trial
there may be imbalances. For example, if the outcome of interest
is customer spend, one arm might have more
customers with high spend in the month
before the experiment starts.
We can improve estimates of the experimental
effect by correcting for such imbalances in variables that are
not affected by the treatment.
This is covariate adjustment (or CUPED, ANCOVA, controls, \etc).

\begin{figure}[ht]
\centering
\includegraphics[width=1.0\textwidth]{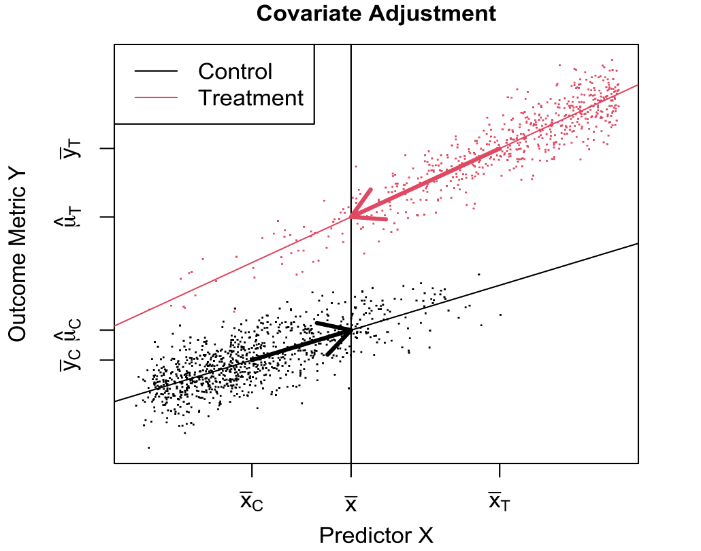}
\caption{Covariate Adjustment corrects for imbalances in predictors between
  control and treatment groups.
  The adjusted estimates $\muhat_C$ and $\muhat_T$ estimate the group means
  {\em if both groups had the same mean for $x$ values.}
  Here the imbalance is exaggerated
  (differences in $x$'s this large are extremely unlikely in
  a randomized controlled trial).
\label{figure:covariateAdjustment}}
\end{figure}

Figure~\ref{figure:covariateAdjustment} shows how this works
for the case of one covariate (one predictor), using linear regression.
The mean $Y$ is clearly larger for the treatment group than
for the control group. However, that is not solely due to the treatment;
the treatment group also has larger $X$ values than the control group,
which inflates $\Ybar_T$ and depresses $\Ybar_C$.
We correct for the imbalance using the predictions at the
common mean $\xbar$.

For multiple regression with $p$ predictors we fit separate regression
models to the control and treatment data, both of the form
\begin{equation}
  \yhat = \betahat_0 + \sum_{j=1}^p \betahat_j x_j
\end{equation}
Then revised estimates for each arm are:
\begin{eqnarray}
  \muhat_C &=& \betahat_{C0} + \sum_{j=1}^p \betahat_{Cj} \xbar_j \nonumber\\
  \muhat_T &=& \betahat_{T0} + \sum_{j=1}^p \betahat_{Tj} \xbar_j
  \label{equation:covariateAdjustmentEstimates}
\end{eqnarray}
where $\xbar_j$ is the common mean of the $j$th predictor.

Fitting separate models is equivalent to fitting a single model
that includes interactions of the treatment variable with all predictors.
We could fit a single model that excludes some interactions;
this corresponds to fitting separate
models but with the constraint that $\betahat_{Cj} = \betahat_{Tj}$
for some values of $j$ (and using the same prediction formulas).

These estimates are a special case of the general rule:
let $\Yhat_{Ci}$
and $\Yhat_{Ti}$ be the control model and treatment model predictions
for observation $i$, then
\begin{eqnarray}
  \muhat_C &=& \meanin \Yhat_{Ci} \nonumber\\
  \muhat_T &=& \meanin \Yhat_{Ti}.
\end{eqnarray}
These averages are over all observations,
regardless of which arm $i$ was assigned to; in other words, we estimate
what the mean responses would be, if both arms had the same distribution
of $x$ values.

\subsection{Standard Errors using Residuals}

To calculate the standard errors for covariate adjusted estimates in
Equation~(\ref{equation:covariateAdjustmentEstimates}),
we begin with the residuals.
For each group (C and T separately), the residual standard deviation is
\begin{equation}
  s^2_r = \frac{1}{n-p-1} \sum (Y_i-\hat Y_i)^2
\end{equation}
Then the standard error is
\begin{equation}
  \SE_{\hat\mu} = \frac{s_r}{\sqrt{n}}
  \label{equation:SEcorrectedMean}.
\end{equation}

We are intentionally excluding a term from this standard error.
Consider the control arm, and let $\sigma^2_C$ be the variance
of residuals relative to the true regression line/plane.
The prediction at $\xbar_C$ is $\Ybar_C$, which has variance $\sigma^2_C/n$,
which we estimate using $s^2_{rC} / n$.
The missing term is the extra variance for predictions at other points,
in particular at $\xbar$.
But in randomized trials with large samples,
$\xbar$ is typically close to the group mean $\xbar_C$,
and the additional variance is negligible.

Similarly, we are not using
heteroskedasticity-consistent (HC) calculations for standard errors
or covariance matrices. HC methods would have a negligible impact
on variances for averages of predictions when $\xbar_c \approx \xbar$.

In fact, our approach avoids the need to ever estimate covariance
matrices for the coefficients. This makes it practical to use
covariates with a large number of levels, \eg customers, using fitting
methods that do not produce covariate matrices.

\section{Covariate Adjustment for Ratio Metrics}
\label{section:ratioAdjustment}

\indent\indent
To apply covariate adjustment to ratio metrics,
we use regression adjustments independently for the numerator $Y$
and denominator $W$, obtaining
\begin{eqnarray}
  \muhat_{YC} = \betahat_{C0} + \sum_{j=1}^p \betahat_{Cj} \xbar_j &&
  \muhat_{YT} = \betahat_{T0} + \sum_{j=1}^p \betahat_{Tj} \xbar_j \nonumber\\
  \muhat_{WC} = \gammahat_{C0} + \sum_{j=1}^p \gammahat_{Cj} \xbar_j &&
  \muhat_{WT} = \gammahat_{T0} + \sum_{j=1}^p \gammahat_{Tj} \xbar_j
\end{eqnarray}
The covariate-adjusted ratio estimates are
\begin{eqnarray}
\thetahat_C &=& \muhat_{YC}/\muhat_{WC} \nonumber\\
\thetahat_T &=& \muhat_{YT}/\muhat_{WT}
\end{eqnarray}

\subsection{Standard Errors using Double Residuals}

We use the delta method to obtain linear approximations for these estimates.
Recall that standard errors for ratio metrics and covariate-adjusted
non-ratio metrics both involve residuals; the standard errors here
involve ``double residuals'' that combine elements of both residuals.
For each arm, let
\begin{equation}
  r_i = y_i - \hat y_i - \thetahat (w_i - \hat w_i)
        \label{equation:doubleResidual}
\end{equation}
These double residuals are like the ratio method residuals in
Equation~(\ref{equation:ratioResiduals}), but with
regression residuals $y_i - \hat y_i$ in place of $y_i$, and
regression residuals $w_i - \hat w_i$ in place of $w_i$.

We calculate the residual variance
\begin{equation}
  s^2_r = \frac{1}{n-p-1} \sum r_i^2
\end{equation}
where the sum is taken across all observations in each arm, $n$ is the number of distinct units of receiving experimental assignment, and $p$ is the number of covariates in the model.

The standard error for the arm is
\begin{equation}
  \SE(\thetahat)
  = \frac{1}{\wbar} \frac{s_r}{\sqrt{n}}
  \label{equation:SEcorrectedRatio}.
\end{equation}

When planning an experiment we use the unadjusted estimate
$\thetahat = \ybar / \wbar$ in Equation~(\ref{equation:doubleResidual}).

\section{Applications}
\label{section:applications}

\indent\indent Here we review the methodology described above,
then consider an example and a meta-analysis.

We began with a review of conventional power analysis methods for
unadjusted differences of means, then described extensions for
(1) clustered data using ratios of means,
(2) ratios of means using the delta method, and
(3) covariate adjustment for both means and ratios of means using regression.
The standard errors of estimates ultimately depend on the
sample standard deviations $s_r$ of certain residuals,
plus division by the estimated denominator mean
($\wbar$ or $\muhat_w$) for ratio estimates.
This is summarized in Table~\ref{table:SEsummmary}.

\begin{table}[h]
\hypertarget{summaryTable}{}
  \centering
\begin{tabular}{rcccc}
     & Estimate & Residual & SE & Equation \\
\hline
  \vspace{-9pt}&&&\\ 
  mean & $\Ybar$
     & $y - \ybar$
     & $s_y / \sqrt{n}$ & \\
  ratio of means & $\Ybar / \Wbar$
     & $y - \thetahat w$
     & $s_r / (\sqrt{n}\wbar)$ & (\ref{equation:SEtheta})\\
  covariate-adjusted & $\muhat_Y$
     & $y - \yhat$
     & $s_r / \sqrt{n}$ & (\ref{equation:SEcorrectedMean}) \\
  cov-adjusted ratio & $\muhat_Y / \muhat_W$
     & $y - \yhat - \thetahat (w - \what)$
     & $s_r / (\sqrt{n}\muhat_w) $ & (\ref{equation:SEcorrectedRatio})\\
\end{tabular}
\caption{Standard errors for means and ratios of means, with and without
  covariate adjustment. These estimates and SEs are for a single arm --- either
  historical data, or one arm in an experiment (with $n$ the sample
  size for the arm).
  For details see the original equations and nearby text.
\label{table:SEsummmary}}
\end{table}

We could use common power analysis tools
by plugging in $s_r/\wbar$, $s_r$, or $s_r/\muhat_W$ in place of $s_y$.

We can incorporate those individual-arm standard errors into
standard errors for the difference between arms, \eg for an adjusted
ratio metric
\begin{equation}
  \SE_{\thetahat_T-\thetahat_C} = \sqrt{\frac{1}{1-\psi} + \frac{1}{\psi}}\frac{s_r}{\muhat_w \sqrt{n}}.
\end{equation}
From there we can calculate the minimum detectable effect
\begin{equation}
  \MDE_{\thetahat_T-\thetahat_C} =  (z_\alpha + z_\beta) \SE_{\thetahat_T-\thetahat_C},
\end{equation}
sample size
\begin{equation}
  n = (z_\alpha + z_\beta)^2 \left(\frac{1}{1-\psi} + \frac{1}{\psi}\right)
  \left(\frac{s_r}{\muhat_w \MDE}\right)^2
\end{equation}
or power
\begin{equation}
  \mbox{power} = 1-\beta = \mbox{CDF}_{\mbox{Normal}}\left(
    \frac{\MDE}{\sqrt{\frac{1}{1-\psi} + \frac{1}{\psi}} \frac{s_{r}}{\muhat_w \sqrt{n}} } - z_\alpha \right)
  \label{equation:summaryPower}.
\end{equation}

\subsection{Example}
\label{section:example}

Let's walk through an application using real-world data. Instacart has store planogram data for some retailers --- detailed descriptions of the exact location of a given product including the aisle number, shelf number, \etc Providing shoppers with this data could speed up their work and increase the proportion of items they find.
This should  make shoppers’ picking experience easier, and let shoppers work through their orders more quickly, increasing the average value of delivered orders as measured by GTV-per-order.

\begin{figure}[h]
\centering
\includegraphics[width=1.0\textwidth]{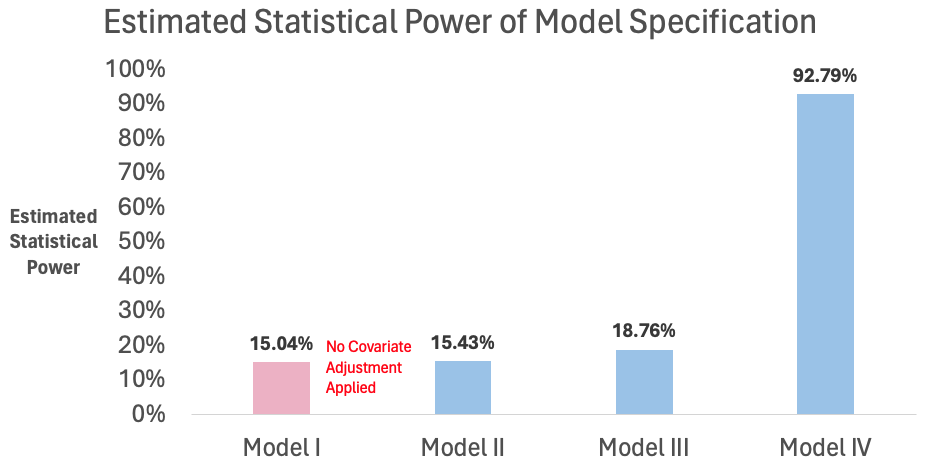}
\caption{Thoughtful application of covariate adjustment can lead to significant improvement in statistical power.
\label{figure:Estimated_Statistical_Power}}
\end{figure}

To test this, consider an A/B test randomized at the shopper level,
with $\alpha = 0.05$ and 10 million orders (or approximately half a million shoppers). We estimate the variance of the average
delivery value from historical data and specify an MDE of \$0.05 per order.
Using standard $t$-tests would give a severely under-powered
experiment, with power of $15\%$ (see Figure~\ref{figure:Estimated_Statistical_Power}).

We can do better using covariate adjustment.
The dollar value of the delivered order $Y$ is highly correlated with the
dollar value of the order the customer placed (see models 3 and 4 in
Table~\ref{table:regressionResults}).
The number of items in the order $W$ is highly correlated with
the sum of estimated probabilities of being in stock (models 2 and 4).
Using either of these predictors alone has minimal value
for covariate adjustment for the ratio of interest, but using them
together gives an $R^2$ for $Y - \thetahat W$ of $0.927$ and
improves power to over $92\%$.

\clearpage

\begin{table}[!htbp]
\centering
{\large \bf{Regression Results: GMV-per-Order}}
\begin{tabular}{c|l|rrrr}
                &               & \multicolumn{4}{c}{Model} \\
  Response      & Predictor     & 1 & 2 & 3 & 4  \\
\hline
  $Y$           & Intercept     &1,434.8 & -107.1 & -2.4 & -6.9 \\
                & Item Availability &    & 98.9   &      & 80.4 \\
                & $Y_{\mbox{pre-fulfillment}}$    &    &        & 0.82 & 0.80 \\
\cline{2-6}
                & \phantom{aaaaaaaaaaaa}R-Squared      & n/a     & 0.872   & 0.992 & 0.992 \\

\hline
  $W$           & Intercept     & 33.7   & 0.187  & 6.43  & 2.31 \\
                & Item Availability &    & 2.13   &       & 2.22 \\
                & $Y_{\mbox{pre-fulfillment}}$    &    &        & 0.015 & -0.0004 \\
\cline{2-6}
                & \phantom{aaaaaaaaaaaa}R-Squared     & n/a & 0.998 & 0.865 & 0.998 \\
\hline
  $Y - \thetahat W$ & \phantom{aaaaaaaaaaaa}R-Squared & n/a & 0.034 & 0.257 & 0.927 \\
\hline
                    & \phantom{aaaaaaaaaaaa}Power     & 15.0\%  & 15.4\%  & 18.8\% & 92.8\%\\
\hline
\end{tabular}
\\
\raggedright
  $Y$ = $\sum$(GMV Amount) \\
  $W$ = \# Orders Fulfilled \\
  Item Availability = Average ML-Estimated Item Availability Score \\
  $Y_{\mbox{pre-fulfillment}}$ = Tentative Chargeable Amount (pre-fulfillment)
  \caption{
Results from regressing shopper-level aggregates of GMV and the number of orders-per-shopper on `Item Availability’ (sum of the mean item availability score as estimated by a ML model), and `$Y_{\mbox{pre-fulfillment}}$' (sum of the tentative chargeable amount). $Y_{\mbox{pre-fulfillment}}$ is highly correlated with Y, and Availability with W. Note: The coefficient estimates are rescaled to avoid disclosing sensitive data. Results are based on 10,000,000 orders and 563,492 shoppers. All $t$-statistics for coefficients are 8.5 or larger.
\label{table:regressionResults}
}
\end{table}

\subsection{Meta Analysis}

We see how covariate adjustment can increase statistical power, but what about our original mandate --- to ship promising treatments as quickly as possible? To explore the impact of covariate adjustment on experiment run times, we conducted a meta-analysis of 3,563 A/B tests comprised of 4,642 individual experiment arms. The response variable for these experiments is Gross Transaction Value (GTV) per customer (this is different from GMV-per-order in Table~\ref{table:regressionResults}). Instacart adjusts for the following covariates: customer GTV measured during the 60-day pre-assignment period, the customer's lifetime value (LTV) as estimated from a machine learning model, and the number of days elapsed since experimental assignment.

Holding statistical power, alpha, and the MDE constant across the covariate adjusted versus non-covariate adjusted versions of these hypothesis tests, we see that the median experiment run time for unadjusted tests is approximately 39 days. In contrast, the median run time using covariate adjustment is 13 days. As a thought exercise, if we were to apply the 26 days of run time saved to all 3,563 experiments, then the total time savings would amount to 253 \textit{years}. In a world where time, is in fact, money, then the value proposition of covariate adjustment is evident.

\section{Concluding Remarks}

\indent\indent
Using regression in the context of a randomized controlled trial provides a straightforward way to perform covariate adjustment.
We must take care not to include covariates that are affected by the treatment, which would bias the results. Nevertheless, covariate adjustment is a powerful tool in our toolkit whenever statistical power is at a premium.

Unfortunately, most off-the-shelf power analysis tools do not support covariate adjustment. These same tools often fail when presented with clustered data and/or ratio metrics. The approach described above provides a way to conduct power analysis in these cases, without the need for complex simulations.

\section{Appendix}
\label{section:appendix}

\indent\indent
Here we discuss issues with standard errors if we view
ratio estimates (\ref{equation:ratioMetric})
as weighted averages of $V$ values, \ie
$\thetahat = \sum_i W_i V_i / \sum_i W_i$.

\begin{figure}[ht]
\centering
\includegraphics[width=1.0\textwidth]{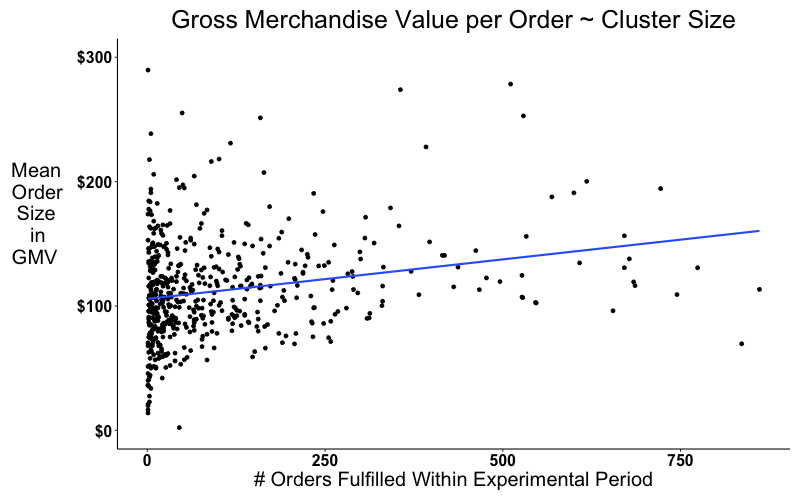}
\caption{Each point corresponds to one shopper; on the $y$ axis is
  $V_i = $ average order size for orders delivered by the shopper,
  and the $x$ axis has number of orders.
  \label{figure:scatterGMV.orders}}
\end{figure}

Consider the two metrics discussed above. GMV-per-order
corresponds to:
\begin{equation}
  V_i = \frac{Y_i}{W_i} = \frac{\sum_j \mbox{GMV}_{ij}}{\mbox{Orders by customer\ }i}
    \label{equation:clusterRatio}
\end{equation}
(shown in Figure~\ref{figure:scatterGMV.orders}),
while the fraction of GMV picked by Instacart shoppers is
\begin{equation}
  V_i = \frac{Y_i}{W_i} = \frac{I(i \mbox{~is Instacart Shopper}) \mbox{GMV}_i}{\mbox{GMV}_i}
  \label{equation:intrinsicRatio}
\end{equation}
where $I(x) = 1$ if $x$ is true, otherwise $0$.

To calculate standard errors for a weighted average
$\thetahat = \frac{\sum_i W_i V_i}{\sum_i W_i}$,
a convenient pair of assumptions would be that the weights
are fixed values and the $V_i$ are independent with
common variance $\sigma^2$; then
$\Var(\thetahat) \frac{\sum_i W_i^2}{(\sum_i W_i)^2} \sigma^2$,
and we could estimate $\sigma^2$ using the sample variance of $V$.

However, neither assumption is reasonable here. The weights
$W_i = n_i$ are the cluster sizes, which are not fixed but instead
depend on which shoppers are randomized into each arm. And
GMV-per-order has higher variability for shoppers with few orders.

Or consider using weighted regression, with a model
\begin{equation}
  V_i = \beta_0 + \beta_1 T
\end{equation}
with weights $W_i = n_i$,
where $T$ is treatment dummy variable.
Then
$\hat\beta_0 = \sum_{i \in C} W_i V_i / \sum_{i \in C} W_i$,
and
$\hat\beta_1 = \sum_{i \in T} W_i V_i / \sum_{i \in T} W_i - \hat\beta_0$
is the estimated experimental effect.
There are many software programs that estimate coefficients for
this weighted regression, and produce standard errors for the
coefficients. Unfortunately, those standard errors would be
incorrect, for the GMV-per-order data in
Figure~\ref{figure:scatterGMV.orders}.
The regression implicitly assumes that the weights
are fixed and unaffected by the treatment, that the variance
of the individual $V$ values is inversely proportional to the
weights, and that the model is correct --- that the expected
value of $V$ depends on treatment arm but not on the number
of orders. The software would estimate residual variance assuming
that model.
Those assumptions are incorrect --- the weights are random,
may be affected by the treatment, the expected value increases with
the number of orders, and the variance decreases more slowly than
suggested (perhaps because more-experienced shoppers handle larger
orders which vary in size more).

In general it is dangerous to rely on canned software to
estimate standard errors for weighted data.
Standard errors depend on the randomization mechanism and how the
weights arise, and the assumptions that software makes may not
hold in your application.




\printbibliography 

\end{document}